\documentclass[aps,prl,showpacs,noshowkeys,amsmath,amssymb,amsfonts,reprint]{revtex4-1}
\DeclareMathOperator{\sign}{sgn}
\usepackage{graphicx}
\usepackage{bm}
\usepackage[colorlinks=true,linkcolor=blue,citecolor=blue,urlcolor=blue]{hyperref}
\usepackage[all]{hypcap}
\begin{document}
\title{Landscape-Inversion Phase Transition in Dipolar Colloids: \\
Tuning the Structure and Dynamics of 2D Crystals}
\author{Ricard Alert}
\email{ricardaz@ecm.ub.edu}
\affiliation{Departament d'Estructura i Constituents de la Mat\`{e}ria, Universitat de Barcelona, Barcelona, Spain}
\author{Jaume Casademunt}
\affiliation{Departament d'Estructura i Constituents de la Mat\`{e}ria, Universitat de Barcelona, Barcelona, Spain}
\author{Pietro Tierno}
\email{ptierno@ub.edu}
\affiliation{Departament d'Estructura i Constituents de la Mat\`{e}ria, Universitat de Barcelona, Barcelona, Spain}
\date{\today}
\begin{abstract}
We study the 2D crystalline phases of paramagnetic colloidal particles with dipolar interactions and constrained on a periodic substrate. Combining theory, simulation, and experiments we demonstrate a new scenario of first-order phase transitions that occurs via a complete inversion of the energy landscape, featuring non-conventional properties that allow for: (i) tuning of crystal symmetry; (ii) control of dynamical properties of different crystalline orders via tuning of their relative stability with an external magnetic field; (iii) an equivalent but independent control of the same dynamic properties via temporal modulations of that field; and (iv) non-standard phase-ordering kinetics involving spontaneous formation of transient metastable domains.
\end{abstract}
\pacs{82.70.Dd, 64.70.pv, 05.70.Fh}
\maketitle
Understanding the structural order of interacting particles above patterned substrates is relevant for many phenomena in condensed matter physics, from atomic adsorption on periodic substrates \cite{Birgeneau1986} to pinning of vortices on arrays of magnetic dots \cite{Martin1997}. It is also of great potential interest for technological applications related to the development of photonic band gap materials \cite{Joannopoulos1995}, chemical sensors \cite{Holtz1997} or antireflection coatings \cite{Min2008}, among others. Main advantages in using colloids as model systems for studies on phase transitions rely on their direct experimental accessibility \cite{Crocker1996,*Anderson2002,*Tan2013} and the possibility of inducing tunable interactions \cite{Yethiraj2007,*Yethiraj2003,*Hynninen2005}. Because of this tunability, several complex structures can be assembled via the application of electric and/or magnetic fields, either static \cite{Wen2000,*Aubry2008,*Leunissen2009,*Demirors2013} or oscillating \cite{Martin2013,*Dobnikar2013,*Martin2003,*Snezhko2005,*Tierno2007,*Osterman2009,*Elsner2009,*Koser2013}.

On a periodic substrate, the equilibrium organization of the particles results from the interplay between pinning, thermal noise, and interparticle interactions. Competition between the inherent symmetries of the substrate and the interactions can produce either intermediate phases \cite{Neuhaus2013} or exotic patterns \cite{Mikhael2008} with exciting perspectives towards the development of new materials. By now, the study of the ordering of colloidal particles on periodic potentials has focused mainly on systems with isotropic interactions, either from electrostatic \cite{Chowdhury1985,*Chakrabarti1995,*Wei1998,*Frey1999,*Zaidouny2013}, magnetic \cite{Mangold2003}, or entropic \cite{Lin2000} origin. Anisotropic interactions have been addressed for non-spherical particle aggregates, namely colloidal molecular crystals \cite{Reichhardt2002,*Brunner2002,*Agra2004,*Sarlah2005,*ElShawish2008,*Reichhardt2009,*ElShawish2011}.

Here we study a system of spherical dipolar particles featuring anisotropic interactions, and confined on top of a two-dimensional striped pattern substrate. We show the existence of two first-order phase transitions between different crystalline orderings as the density of particles is changed and as an external magnetic field is applied. The latter phase transition proceeds via a new mechanism based on a global inversion of the energy landscape, where fixed equilibrium states exchange their local stable/unstable character, in contrast to the exchange of global stability/metastability characteristic of usual first order transitions. We term this phenomenon ``landscape-inversion phase transition" (hereinafter LIPT). In addition, we show that it is possible to induce a shift of this transition by driving the system via a time-modulation of the magnetic field. This effect is due to the nonlinear contribution of the magnetic field to the energy, and enables an unstable equilibrium crystalline phase to be dynamically stabilized by a zero-average driving. Periodic forcings have been recently used to drive colloidal systems towards equilibrium crystals avoiding kinetically arrested phases \cite{Swan2014}. In contrast, here we employ these drivings to stabilize an unstable equilibrium ordering.

\begin{figure}[!htbp]
\includegraphics[width=\columnwidth]{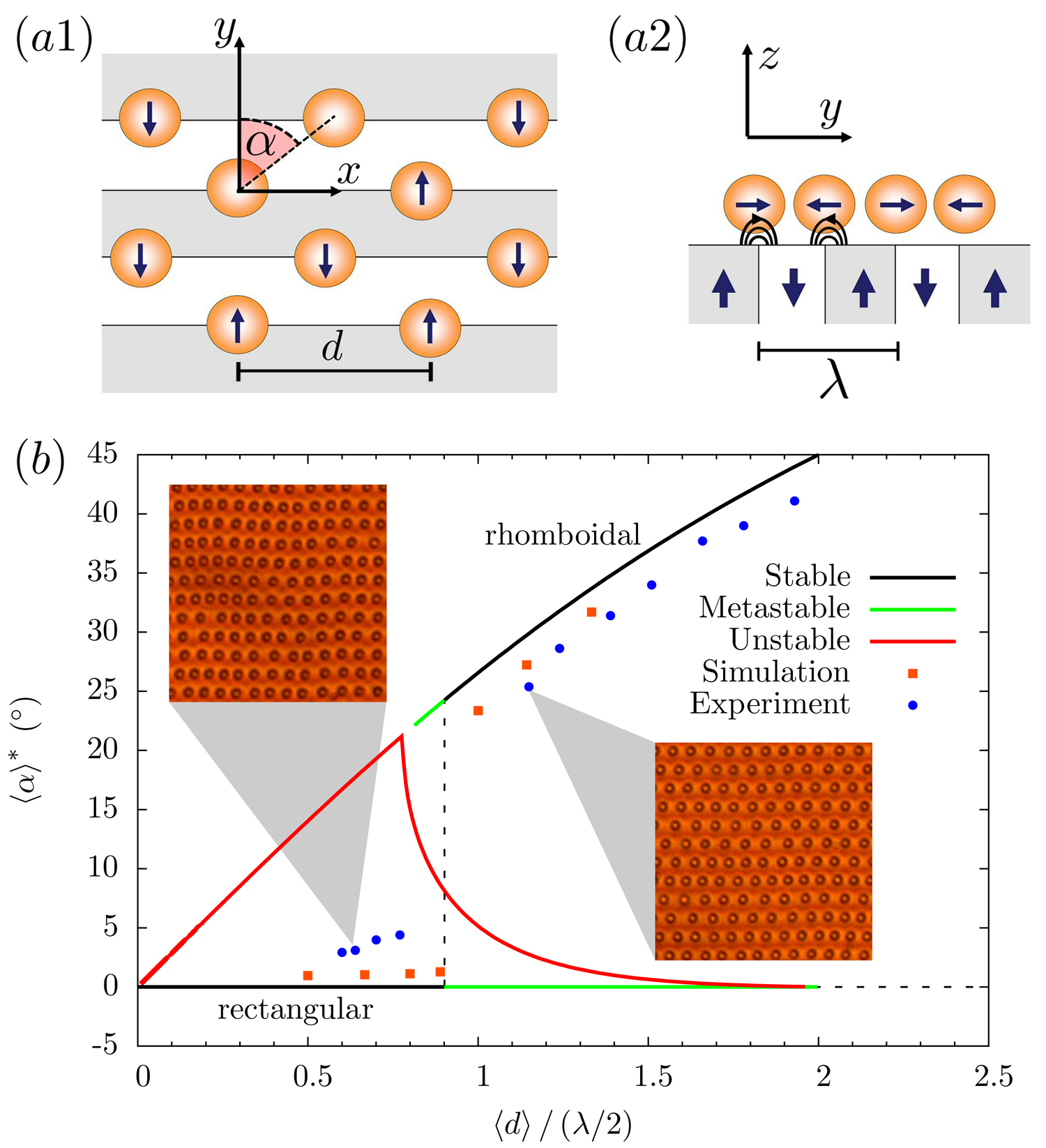}
\caption{\label{fig1}(Color online) (a)\label{fig1a} Schematics of the system showing the striped substrate and the paramagnetic colloids with induced dipolar moments from top (a1)\label{fig1a1} and side (a2)\label{fig1a2} views. (b)\label{fig1b} Structural phase diagram, showing the equilibrium positional angle $\left\langle\alpha\right\rangle^*$ as a function of the inverse linear density $\left\langle d\right\rangle$ along the lines. Theoretical prediction (black), experimental (blue circles), and simulation (orange squares) results are shown for the equilibrium stable state. A discontinuous phase transition from a rectangular to a rhomboidal phase occurs when the density $\rho=1/\left\langle d\right\rangle$ is decreased (dashed black). Predictions for the metastable (green) and unstable (red) states are also plotted. Insets display small overviews $(64\times 48\,\mu$m$^2)$ of the experimental system.}
\end{figure}

In our system, the one-dimensional periodic potential of the substrate is generated by a uniaxial ferrite garnet film (FGF) grown by liquid phase epitaxy \cite{Tierno2009}. The FGF features a series of parallel striped domains with periodicity $\lambda=2.6\,\mu$m and opposite magnetization between consecutive ones, separated by Bloch walls (BWs), i.e. narrow transition regions with maximum stray magnetic field $H_{\text{stray}}=1.3\times 10^4\,$A/m. An aqueous suspension of paramagnetic colloids with diameter $2a=1\,\mu$m (Dynabeads Myone) is deposited on top of the FGF and, after sedimentation, particles locate above the BWs, acquiring a magnetic moment $\vec{m}= V\chi\vec{H}_{\text{stray}}$ (Fig. \ref{fig1a}(a)), with $V$ the particle volume and $\chi\sim 1$ the magnetic volume susceptibility. Above the BWs, particles display a small lateral diffusion, with $D_x=\left(7.0\pm0.4\right)\times 10^{-3}\,\mu$m$^2$/s, while motion in the direction transversal to the BWs is suppressed. As shown in Fig. \ref{fig1a}(a), the magnetic landscape of the substrate arranges the induced dipoles in an antiferromagnetic order between lines, with moments oriented parallel (antiparallel) for particles located along the same BW (consecutive BWs). Video microscopy is used to track particles recorded for $\sim 5\,\text{min}$ at a rate of $60\,$Hz on an observation area of $140\times 105\,\mu$m$^2$.

As an order parameter for the crystalline phases we use the spatial average $\left\langle\alpha\right\rangle$ of the positional orientation angle $\alpha$ between nearest neighbor particles located on adjacent stripes (Fig. \ref{fig1a1}(a1)), with equilibrium value denoted by $\left\langle\alpha\right\rangle^*$. The fact that the pattern periodicity $\lambda$ is fixed in our system imposes a geometrical constraint that allows a description of the crystalline structure in terms only of the angle $\left\langle\alpha\right\rangle$. For homogeneous states, an explicit form of the energy landscape of the system can be derived directly from the dipolar interactions between nearest neighbors \footnote{Numerical simulations show the accuracy of the nearest neighbors approximation for the field strengths used here.} as a function of the stray field of the substrate, $H_{\text{stray}}$, the average distance between particles in a line $\left\langle d\right\rangle$ (the inverse of the longitudinal linear density $\rho=1/\left\langle d\right\rangle$), and $\left\langle\alpha\right\rangle$ (see details in \cite{Alert}). Note that the linear density along the direction transversal to the lines is fixed by the constraint of the substrate pattern, and only the linear density of particles along each line, $\rho$, is changed. The equilibrium configuration given by energy minimization is independent of $H_{\text{stray}}$ but depends on $\rho$, i.e. on $\left\langle d\right\rangle$. Fig. \ref{fig1b}(b) shows the theoretical prediction for the phase diagram (black lines) and both the experimental and the numerical verification, the latter including thermal fluctuations explicitly. The slight systematic deviation of data from the theory can be attributed to the neglected fluctuations in the mean-field theoretical approach (see discussion in \cite{Alert}). The theoretical model also predicts the metastable and unstable states, as shown in Fig. \ref{fig1b}(b).

An important feature of the phase diagram in Fig. \ref{fig1b}(b) is the presence of a discontinuous phase transition from a rectangular ordering at high densities to a rhomboidal one at low densities. The transition is predicted to occur at $\left\langle d\right\rangle_t=0.893(8)\lambda/2$. The hysteresis region is also reproduced in simulations \cite{Alert}.
This phase transition results from the anisotropy of the dipolar magnetic interactions. These interactions 
between dipoles in consecutive BWs can be either repulsive or attractive depending on the positional angle $\alpha$. For high (low) densities, the geometrical constraints favor the attractive (repulsive) interaction across stripes and the particles arrange in a rectangular (rhomboidal) order. The role of the density in colloids on periodic substrates was experimentally studied with regard to the melting transition \cite{Bechinger2001}. 
Also, a density-induced phase transition between crystalline orderings has been recently predicted in a system with isotropic interactions \cite{Neuhaus2013}. Here, we predict and experimentally demonstrate a similar phase transition but arising from the anisotropy of the interactions. In addition, the crystal symmetry of the rhomboidal phase can be continuously tuned by changing the density.

Our system displays another phase transition when an external magnetic field is applied in the plane of the substrate, perpendicular to the BWs (Fig. \ref{fig1a}(a)), $\vec{H}=H\hat{y}$. Such an in-plane magnetic field enforces (weakens) the dipoles with moments parallel (antiparallel) to it. 
For $H > H_{\text{stray}}$ the field flips the anti-parallel moments, forcing the 
system to transit to a ferromagnetic phase, as opposed to the original antiferromagnetic-like phase.
\begin{figure}[!htbp]
\includegraphics[width=\columnwidth]{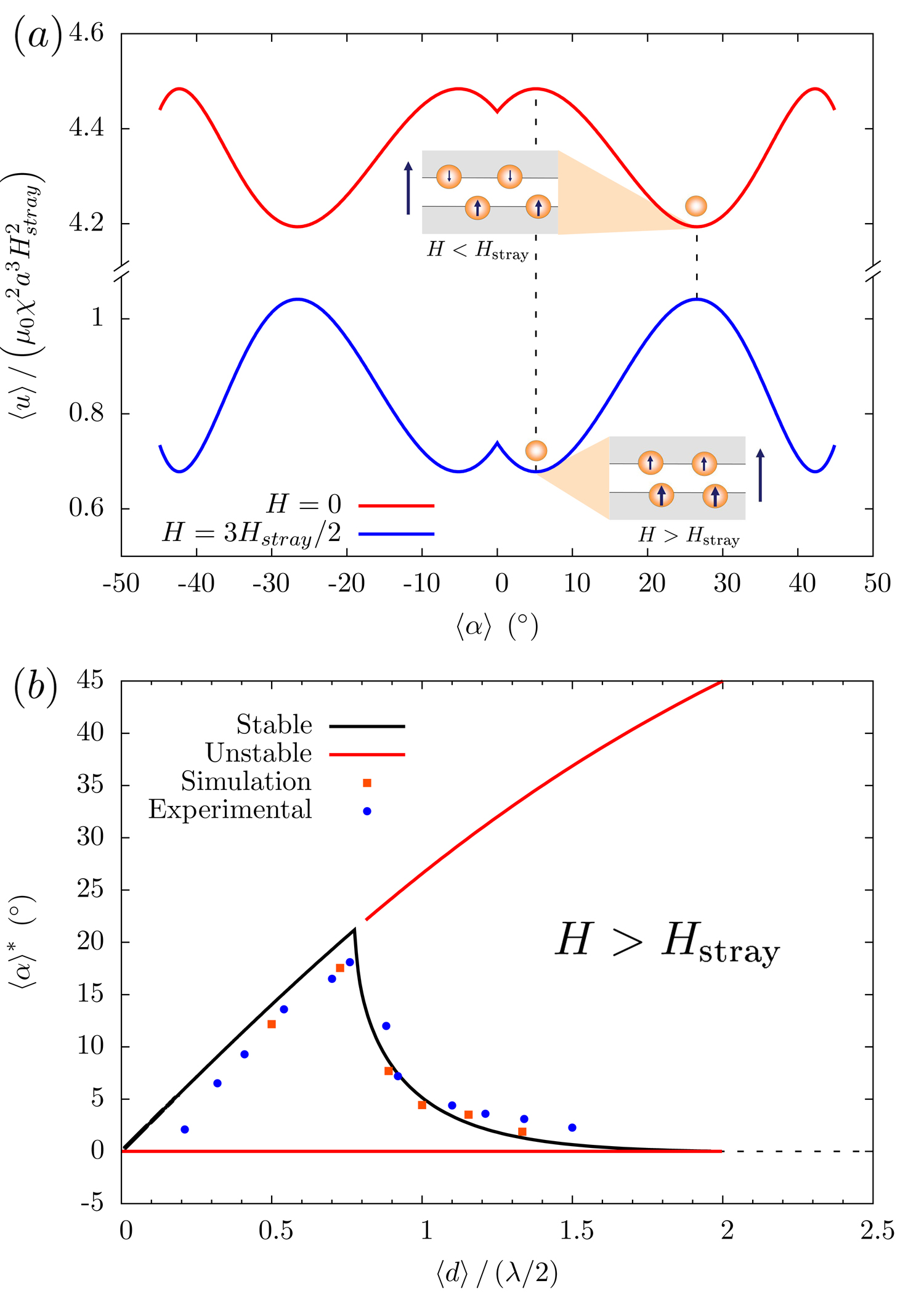}
\caption{\label{fig2}(Color online) Landscape-inversion phase transition (LIPT). (a)\label{fig2a} Mean-field energy as a function of the order parameter $\left\langle\alpha\right\rangle$ for a density corresponding to $\left\langle d\right\rangle=\lambda/2$ and external magnetic field below (red) and above (blue) the substrate field $H_{\text{stray}}$. The anisotropic part of the energy landscape completely flips for $H>H_{\text{stray}}$, thus inducing the LIPT. Insets display the corresponding equilibrium orderings. (b)\label{fig2b} Phase diagram for $H>H_{\text{stray}}$, complementary in terms of stability to that of Fig. \ref{fig1b}(b). Theoretical predictions (black) are shown along with experimental (blue circles) and simulation (orange squares) results obtained for $H=2\times 10^4\,$A/m and $H=5\times 10^4\,$A/m, respectively. Predicted unstable states (red) are also indicated.}
\end{figure}

The external magnetic field can be incorporated into the model, rendering an average energy per particle
\begin{equation} \label{eq nonlinear}
\left\langle u\right\rangle\sim \left(H_{\text{stray}}^2-H^2\right)f\left(\left\langle d\right\rangle,\left\langle\alpha\right\rangle\right)
\end{equation}
plus other isotropic terms, where $f\left(\left\langle d\right\rangle,\left\langle\alpha\right\rangle\right)$ is the same function that gives rise to the phase diagram of Fig. \ref{fig1b}(b) \cite{Alert}. The total energy $\left\langle u\right\rangle$ is plotted in Fig. \ref{fig2a}(a) as a function of the order parameter $\left\langle\alpha\right\rangle$, for a given density and for values of the external field below and above the substrate field. The effect of the factor $\left(H_{\text{stray}}^2-H^2\right)$ is to completely flip the anisotropic (angular-dependent) part of the energy for $H>H_{\text{stray}}$, while the angular function $f\left(\left\langle d\right\rangle,\left\langle\alpha\right\rangle\right)$ remains unaffected. The value $H=H_{\text{stray}}$  defines the LIPT, where equilibrium stable states become unstable and viceversa, but remaining at the same values of the order parameter. Therefore, the phase diagram under these conditions, as shown in Fig. \ref{fig2b}(b), has the same shape as the one in Fig. \ref{fig1b}(b), having exchanged the stability of the equilibrium states corresponding to the cases $H>H_{\text{stray}}$ and $H<H_{\text{stray}}$. Even when not crossing $H_{\text{stray}}$, variations of the external field modify the relative stability of the different phases while preserving the equilibrium  value of the order parameter. Therefore, dynamical properties of these crystalline states, such as fluctuations or relaxation rates, can be externally tuned.

\begin{figure}[!t]
\includegraphics[width=\columnwidth]{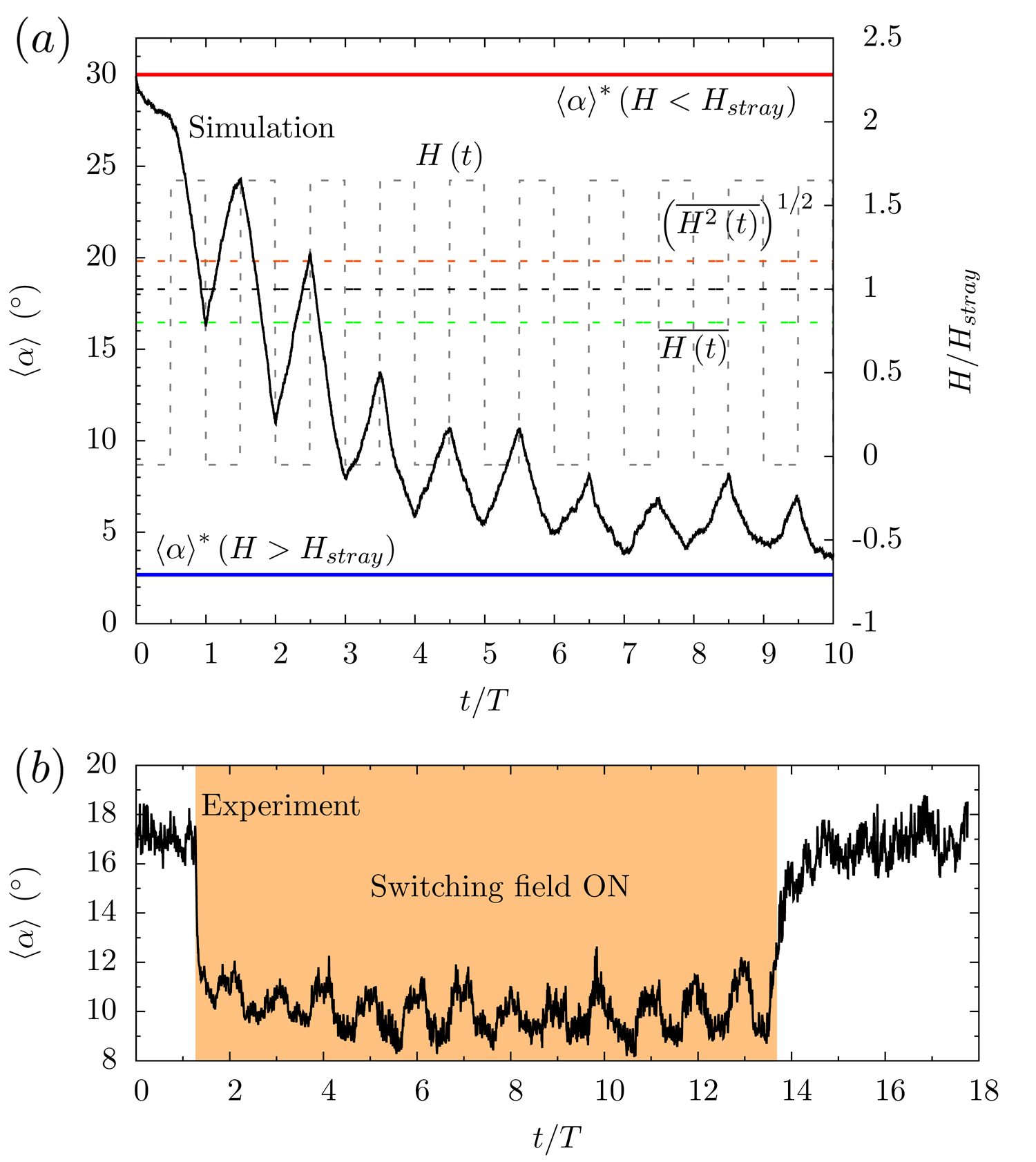}
\caption{\label{fig3}(Color online) Dynamical stabilization of the unstable crystalline ordering. (a)\label{fig3a} Simulated temporal evolution of the order parameter $\left\langle\alpha\right\rangle$ from the low-field phase (red) to the high-field phase (blue) even if the average field corresponds to the low-field phase. The right axis shows the external field driving $\overline{H\left(t\right)}=0.8H_{\text{stray}}$, $\Delta H=1.7H_{\text{stray}}$, $T=1/\nu=0.05\,$s (dashed grey), with an average (dashed green) below the transition threshold in static conditions (dashed black). A nonlinearity makes the system respond to an effective static value (dashed orange) above the threshold. (b)\label{fig3b} Experimental demonstration of the dynamical stabilization taking place when the temporal modulation is applied (orange shading). Quantitative discrepancies stem from more pronounced spatial inhomogeneities in the experiment.}
\end{figure}

Next we explore the LIPT when the system is driven by temporal oscillations of the external magnetic field, so that a given configuration alternates between stable and unstable dynamics.  Note that the only role of the density is to set the two equilibrium configurations between which the system will transit. In particular, we fix an inverse density $\left\langle d\right\rangle=\lambda/\sqrt{3}$ corresponding to the triangular phase $\left\langle\alpha\right\rangle^*=30^\circ$ for low magnetic fields ($H<H_{\text{stray}}$, Fig. \ref{fig1b}(b)), and to a rhomboidal phase $\left\langle\alpha\right\rangle^*=2.7^\circ$ for high magnetic fields ($H>H_{\text{stray}}$, Fig. \ref{fig2b}(b)). We take a square-wave modulation of frequency $\nu$ and peak-to-peak amplitude $\Delta H$ on the external field, $H\left(t\right)=H_0+\left(\Delta H/2\right)\sign\left(\sin\left(2\pi\nu t\right)\right)$, with an offset $H_0$ corresponding to the temporal average $\overline{H\left(t\right)}=H_0$. We find that, for a certain range of $\Delta H$ and sufficiently high $\nu$, the system can be dynamically stabilized at the high-field phase even though this phase is unstable for the average field, $\overline{H\left(t\right)}<H_{\text{stray}}$. Similarly, the energetically stable phase will be dynamically destabilized \footnote{The effect is robust to changes in the shape of the modulation. In particular we have checked the effect explicitly for a Gaussian white noise signal for the applied field.}. We confirm this phenomenon via both numerical simulations (Fig. \ref{fig3a}(a)) and experiments (Fig. \ref{fig3b}(b)); see video in \cite{Alert}. 

\begin{figure}[!t]
\includegraphics[width=\columnwidth]{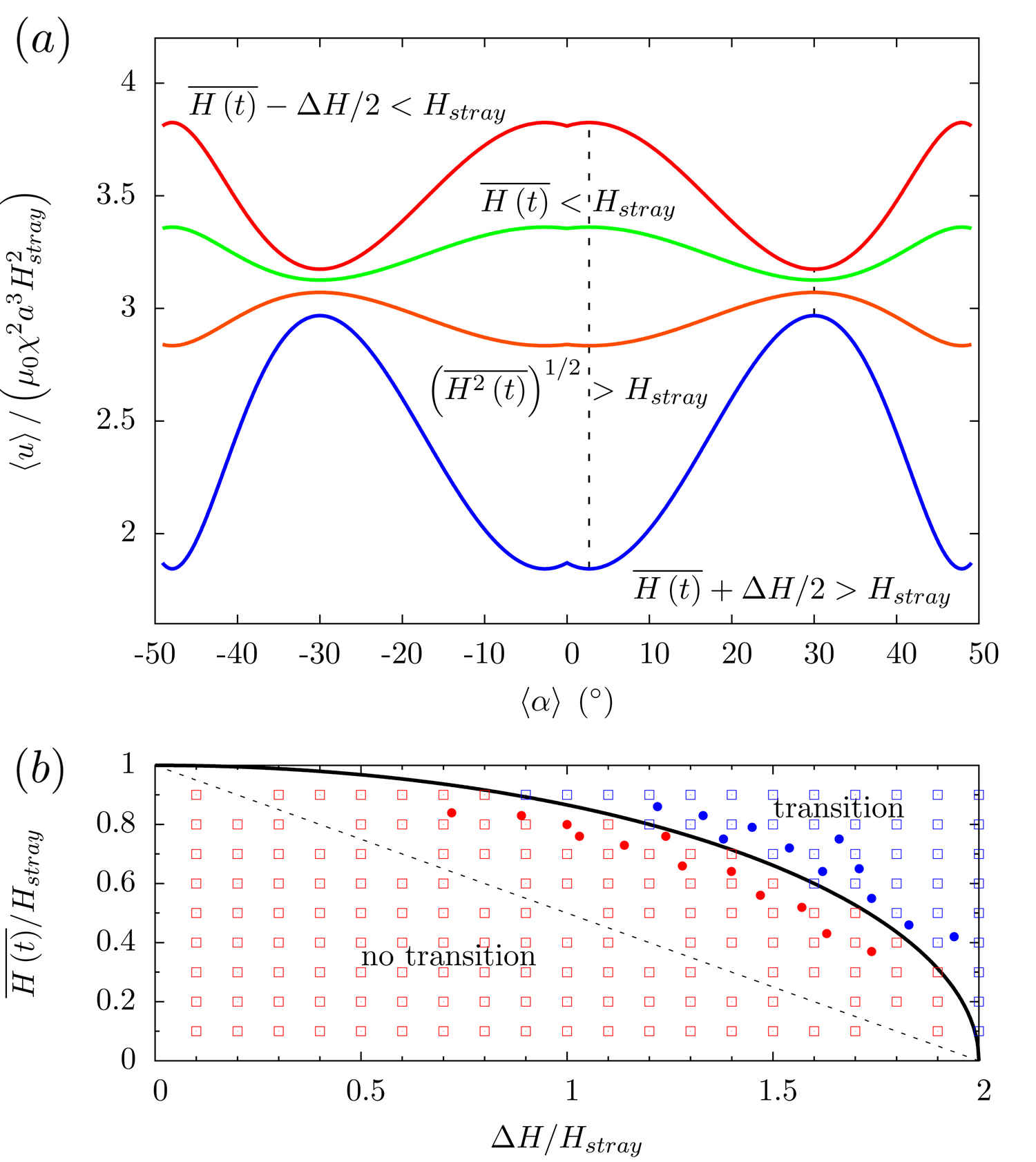}
\caption{\label{fig4}(Color online) Conditions for the dynamically induced LIPT. (a)\label{fig4a} Energy landscapes for the low-field (red) and high-field (blue) semiperiods. When rapidly switched between both states, a linear response system would respond to the energy corresponding to the average field (green), as opposed to the actual temporal average energy profile (orange) in our nonlinear system. The plotted curves correspond to $\overline{H\left(t\right)}=0.8H_{\text{stray}}$, $\Delta H=1.7H_{\text{stray}}$, a case for which the system is dynamically stabilized in the equilibrium unstable state. (b)\label{fig4b} Stationary state diagram showing the presence (blue) or absence (red) of the dynamically induced transition, as a function of the external field only. The theoretical border (black) is verified in experiments (circles) and simulations (open squares). $\overline{H\left(t\right)}+\Delta H/2=H_{\text{stray}}$ (dashed) is a singular case because the moments of the particles in one of every two rows exactly vanish due to the external field.}
\end{figure}

The phenomenon can be understood by considering the case of fast driving, for which the system responds to the temporal average of the energy landscape. Fig. \ref{fig4a}(a) shows examples of the flipping energy landscapes. Due to the nonlinear coupling of the magnetic field in the energy function, the average energy landscape is not given by $\overline{H\left(t\right)}$ but by $(\overline{H^2\left(t\right)})^{1/2}$, which is always larger and may eventually cross the transition boundary. In this case, the high-field phase is stabilized, even though the value $\overline{H\left(t\right)}$ would correspond to the low-field phase. Hence, the condition for this to occur reads $(\overline{H^2\left(t\right)})^{1/2}>H_{\text{stray}}>\overline{H\left(t\right)}$, as illustrated in Fig. \ref{fig4a}(a). In parallel to the static case, this corresponds to a flip in the anisotropic part of the \emph{temporal average} of the energy landscape. For a square signal, the condition can be rewritten as $\left(\Delta H\right)^2>4(H_{\text{stray}}^2-\overline{H\left(t\right)}^2)$, which is plotted in Fig. \ref{fig4b}(b) together with simulation and experimental data verifying it. Remarkably, this is a condition only over the external field and does not involve the density. Therefore, this provides an alternative way to tune the relative stability and dynamics of the crystalline phases. For low frequencies, the system follows the external field and oscillates between the two states rather than becoming dynamically stabilized. The role of the driving frequency is further discussed in \cite{Alert}.

The shift of the onset of instability via temporal drivings has been theoretically studied both for specific systems \cite{Onuki1982b,Torrent1988} and general models \cite{Lucke1985,*DePasquale1984,*Becker1994}, and also observed in some experiments on binary fluid mixtures \cite{Joshua1983}, Rayleigh-B\'{e}nard convection \cite{Ahlers1984}, and water vortex patterns \cite{Lauret2013}. In contrast, here we demonstrate that an equilibrium-unstable crystalline phase can be stabilized by a temporal modulation. Moreover, while shifts of instability thresholds can usually be traced down to a nonlinear or multiplicative coupling of the time-modulated parameters, in the case of a LIPT a fully dynamical picture can be constructed. Within the mean-field approach, this allows for the determination of the actual relaxation time towards the newly stabilized phase, as well as the escape time from the destabilized one. As discussed in \cite{Alert}, the exponential escape (relaxation) rate from (to) the destabilized (stabilized) phase takes the form $g\left(\left\langle d\right\rangle,\left\langle\alpha\right\rangle^*\right)(\overline{H^2\left(t\right)}-H_{\text{stray}}^2)$, where the function $g$ can be determined from the expression of the energy \cite{Alert}.

Finally we discuss the new features introduced by the LIPT scenario in phase-ordering kinetics. A quench inverting the energy landscape naturally leaves the system at an unstable state (see Fig. \ref{fig2}(a)). Its early relaxation will then form interfaces separating domains of two locally stable phases. In contrast to standard phase-ordering kinetics \cite{Bray1994}, the two coexisting phases will generically have different relative stability. Half of the system will thus evolve towards a metastable phase that was not present in the initial condition. The dynamical generation of metastable domains was theoretically predicted via a more complex mechanism \cite{Bechhoefer1991}, while it arises naturally within the LIPT scenario. Metastable domains will subsequently be invaded by the stable phase at a finite speed (slightly corrected by curvature) \cite{VanSaarloos2003}. Hence, the phase-ordering kinetics is not curvature-driven but governed by a front propagation mechanism, which will break the usual self-similar coarsening. Note that varying the magnetic field changes the relative stability of the coexisting phases, and hence the corresponding front speed, so that the phase-ordering kinetics can also be externally tuned. Furthermore, from the discussion above one can expect an equivalent control of the phase-ordering kinetics with fast modulations of the magnetic field.

In summary, we have shown that magnetic colloids on a periodic substrate can display novel structural phase transitions due to the anisotropy of their interactions. Specifically, a new phase transition scenario involving a full reversal of the energy landscape has been demonstrated. It exhibits remarkable versatility to external control by a magnetic field, that can tune and invert the relative stability of the different phases, and hence modify their dynamics (relaxation properties, front propagation, fluctuations, etc), without modifying their crystalline order. In turn, the latter can be tuned continuously with the density variable. Moreover, a nonlinear coupling of the external field allows one to use fast temporal modulations of this field to achieve the same degree of control of statics and dynamics of the different phases, regardless of the behavior corresponding to the average value of the field. These phenomena are theoretically understood, thus opening the possibility to engineer or look for other physical systems exhibiting similar behavior, and explore possible practical applications. From a fundamental point of view, the non-standard phase-ordering kinetics is a particularly appealing open problem.

\begin{acknowledgments}
We acknowledge T. H. Johansen for providing the FGF and J. Dobnikar for stimulating discussions. R.A. acknowledges support from Fundaci\'{o} ``la Caixa'', and University of Cambridge under the EU project ITN-COMPLOIDS 234810. R.A. and P.T. acknowledge support from FIS2011-15948-E. P.T. acknowledges support from the ERC project 335040 and from the program RYC-2011-07605. J.C. acknowledges support from MICINN under project FIS2010-21924-C02-02 and Generalitat de Catalunya under project 2009-SGR-14.
\end{acknowledgments}
\bibliography{bibliographyabbreviated}
\end{document}